\documentclass[twocolumn,superscriptaddress,prl,showpacs,letter]{revtex4}

\usepackage{amssymb}
\usepackage{graphicx}
\usepackage{natbib}
\usepackage{multirow}
\usepackage{amsmath}
\usepackage{times}

\begin{document}

\title{Electron spin dynamics and electron spin resonance in graphene}

\author{Ferenc Simon}
\email{ferenc.simon@univie.ac.at}
\affiliation{Budapest University of Technology and Economics, Institute
of Physics and Condensed Matter Research Group of the Hungarian Academy of
Sciences, H-1521 Budapest, Hungary}
\affiliation{Fakult\"{a}t f\"{u}r Physik, Universit\"{a}t Wien, Strudlhofgasse 4, 1090 Wien, Austria}

\author{Ferenc Mur\'{a}nyi}
\affiliation{Physik-Institut der Universit\"{a}t Z\"{u}rich, Winterthurerstrasse 190, CH-8057
Z\"{u}rich, Switzerland}

\author{Bal\'{a}zs D\'{o}ra}
\affiliation{Budapest University of Technology and Economics, Institute
of Physics and Condensed Matter Research Group of the Hungarian Academy of
Sciences, H-1521 Budapest, Hungary}

\begin{abstract}
A theory of spin relaxation in graphene including intrinsic, Bychkov-Rashba, and ripple spin-orbit coupling is presented.
We find from spin relaxation data by Tombros \textit{et al.} [Nature \textbf{448}, 571 (2007).] that intrinsic spin-orbit coupling dominates over other contributions with a coupling constant of 3.7 meV. Although it is 1-3 orders of magnitude larger than those obtained from first principles, we show that comparable values are found for other honeycomb systems, MgB$_2$ and LiC$_6$; the latter is studied herein by electron spin resonance (ESR). We predict that spin coherence is longer preserved for spins perpendicular to the graphene plane, which is beneficial for spintronics. We identify experimental conditions when bulk ESR is realizable on graphene.
\end{abstract}


\maketitle


\textit{Introduction}. The discovery of graphene \cite{NovoselovSCI2004} stimulated enormous interest due its fundamentally and technologically important properties. One potential application is in spintronics \cite{FabianRMP}, i.e. when the electron spin degree of freedom is utilized as information carrier. The principal parameter governing spintronic usability is the spin relaxation time (also referred to as spin-lattice relaxation time), $\tau_{\text{s}}$, which characterizes how an injected non-thermal equilibrium spin state decays. For realistic applications, $\tau_{\text{s}}$ longer than 10-100 ns is required. A general, often cited concept is that "pure materials made of light elements" can reach this limit. The huge mobility of charge carriers in graphene (approaching $10^6\,\text{cm}^2\text{/Vs}$ \cite{BolotinHighMobility}), the light nature of carbon, and the low-dimensionality of this material are the reasons for the high expectations for its spintronic applications. This is supported by the long spin relaxation time in light metals such as e.g. Li \cite{BeuneuMonodPRB1978} or in low-dimensional conductors \cite{ForroPRB1987}.

Therefore it came as a surprise that $\tau_{\text{s}}$ as short as 60-150 ps are observed in spin transport experiments on graphene \cite{JozsaNat, JozsaPRL2008}, which renders it unusable for such applications. The understanding of this experimental result is therefore of great importance. Theory of spin relaxation are split into two different classes: materials with inversion symmetry (e.g. Na or Si) and to materials where the inversion symmetry is broken either in the bulk (e.g. III-V semiconductors such as GaAs) or in two-dimensional heterostructures. The Elliott-Yafet (EY) theory \cite{Elliott,YafetReview} explains the former case, where only intrinsic (i.e. atomic) spin-orbit coupling (SOC) is present, $L_\text{i}$, and predicts that spin ($\Gamma_{\text{s}}=\hbar/\tau_{\text{s}}$) and momentum relaxation rates ($\Gamma=\hbar/\tau$, $\tau$ is the momentum relaxation time) are proportional: $\Gamma_\text{s}=\alpha_{\text{i}}\frac{L_\text{i}^2}{\Delta^2}\Gamma$. Here $\alpha_{\text{i}}=1..10$ is band structure dependent \cite{BeuneuMonodPRB1978}, $\Delta$ is the energy separation of a neighboring and the conduction band.

The relaxation for broken inversion symmetry is explained by the Dyakonov-Perel (DyP) theory. It applies either when the symmetry breaking is in the bulk, (the Dresselhaus SOC \cite{DresselhausPR1955}, $L_{\text{D}}$) or when it happens for a heterolayer structure (the Bychkov-Rashba SOC \cite{BychkovRashba1,BychkovRashba2}, $L_{\text{BR}}$). The DyP theory shows that the spin and momentum relaxation rates are inversely proportional: $\Gamma_\text{s}=\alpha_{\text{D/BR}}L_{\text{D/BR}}^2/\Gamma$, where $\alpha_{\text{D/BR}}\approx \,1$.

A link between the EY and the DyP was found recently \cite{DoraSimonStronglyCorrPRL2009}: for metals with inversion symmetry but rapid momentum scattering, the generalization of the EY theory leads to $\Gamma_\text{s}=\alpha_{\text{i}}\frac{L_\text{i}^2}{\Delta^2+\Gamma^2}\Gamma$, which gives a DyP like spin relaxation when $\Gamma > \Delta$.

Three sources of SOC are present in graphene: intrinsic, BR type (due to the symmetry breaking by a perpendicular electric field), and the ripple related (which is due to the inevitable ripples in graphene). However, the role and magnitude of these SOC parameters is a debated issue. Estimates for the intrinsic SOC ranges two orders of magnitude; 0.9-200 $\mu\text{eV}$ \cite{HuertasPRB2006,FabianPRB2009,KaneMelePRL2005}, whereas value of the BR SOC appears to be settled to 10-36 $\mu\text{eV}$ per V/nm (Refs. \cite{FabianPRB2009} and \cite{HuertasPRB2006}, respectively). The effect of the substrate for the spin relaxation is also unsettled \cite{FabianCM0905.0424}. Given this debate, a description is required which enables comparison with the spin transport data.

Here, we present the theory of spin relaxation in graphene including intrinsic, BR, and ripple spin-orbit coupling. We analyze the spin transport data from Refs. \cite{JozsaNat,JozsaPRL2008,JozsaPRB2009} and we find that the intrinsic SOC dominates the relaxation with a large, unexpected magnitude.
We discuss two similar honeycomb systems; MgB$_2$ and LiC$_6$, and show that they exhibit similar intrinsic SOC. The result predicts a strong anisotropy of the spin relaxation time. We study the feasibility of bulk electron spin resonance (ESR) spectroscopy on graphene and pinpoint experimental conditions when it is possible. ESR would allow a direct, spectroscopic measurement of $\tau_{\text{s}}$ (Ref. \cite{KipKittelPR1952}), which underlines its importance \cite{ForroPSSB2009}.

\textit{Experimental}. We prepared Li intercalated HOPG graphite by the "immersion into molten Li" method \cite{FischerLiC6}. The golden color of the samples attested the LiC$_6$ intercalation level \cite{DresselhausAP2002}. Freshly cleaved samples were sealed under He in quartz tubes for the ESR experiment.

\textit{Spin relaxation in graphene}. Low energy excitations around the $K$ point of the Brillouin zone are described by a two-dimensional Dirac equation:
\begin{equation}
H=v_{\text{F}}(\sigma_xp_x+\sigma_yp_y),
\end{equation}
with the $v_{\text{F}} \approx 10^6$~m/s Fermi velocity \cite{NovoselovSCI2004}.
The spin-orbit interaction in graphene is given by \cite{HuertasPRB2006}:
\begin{equation}
H_{\text{SO}}=L_{\text{i}}\sigma_zS_z+\frac{L_{\text{BR}}+L_{\text{ripple}}({\bf r})}{2}(\sigma_xS_y-\sigma_yS_x),
\end{equation}
where $L_{\text{i}}$, $L_{\text{BR}}$, and $L_{\text{ripple}}$ are the SOC's of the intrinsic, BR, and ripple terms, respectively. $L_{\text{ripple}}(\bf r)$ is Gaussian correlated random variable, $\langle L_{\text{ripple}}({\bf r})L_{\text{ripple}}({\bf r}')\rangle\sim \delta({\bf r-r}')$.

The spin relaxation rates induced by these SOC's are additive in lowest order provided max$(L_{\text{i}},L_{\text{BR}},L_{\text{ripple}})\ll$max$(\Gamma,\mu)$:

\begin{gather}
\Gamma_{\text{s}}=\Gamma_{\text{s},\text{i}}+\Gamma_{\text{s},\text{BR}}+\Gamma_{\text{s},\text{ripple}}\
\end{gather}

\noindent The contributions from the intrinsic ($\Gamma_{s,\text{i}}$), BR ($\Gamma_{s,\text{BR}}$), and ripple ($\Gamma_{s,\text{ripple}}$) relaxation rates are obtained using the Mori-Kawasaki formula similar to that used in Ref. \cite{SimonPRL2008} considering the conical band structure and the $K,\, K'$ degeneracy:

\begin{gather}
\Gamma_{\text{s},\text{i}}=\delta_{\nu,\parallel}\frac{L_{\text{i}}^2\arctan (\mu/\Gamma)}{2\pi \mu\cdot\widetilde{\mu}(\mu,\Gamma)}\Gamma,\label{IntrinsicRelaxation}\\
\Gamma_{\text{s},\text{BR}}=\frac{\left(2\delta_{\nu,\perp}+\delta_{\nu,\parallel}\right)L_{\text{BR}}^2}{16\pi\widetilde{\mu}(\mu,\Gamma)}\left[1+\left(\frac{\mu}{\Gamma}+\frac\Gamma\mu\right)\arctan\left(\frac{\mu}{\Gamma}\right)\right],\label{RashbaRelaxation}\\
\Gamma_{\text{s},\text{ripple}}=\frac{\left(2\delta_{\nu,\perp}+\delta_{\nu,\parallel}\right)\pi}{32}L_{\text{ripple}}^2\rho(\mu,\Gamma)\label{RippleRelaxation}
\end{gather}

\noindent $\nu=\parallel,\,\text{or}\,\perp$ is the spin polarization direction with respect to the graphene plane; e.g. $\nu=\parallel$ in the spin transport experiments \cite{JozsaNat}.
Here, $\mu$ is the chemical potential and $\widetilde{\mu}(\mu,\Gamma)=-\frac{\Gamma}{\pi} \ln \left(\frac{\mu^2+\Gamma^2}{D^2} \right)+|\mu|\left( 1-\frac{2}{\pi}\arctan \left( \frac{\Gamma}{|\mu|} \right)\right)$ is the
pseudo chemical potential ($D\approx 3$ eV is the cutoff in the continuum theory) which appears in the expression of the density of states (DOS), $\rho(\mu,\Gamma)$, with finite $\mu$ and $\Gamma$:

\begin{gather}
\rho(\mu,\Gamma)=\frac{2 A_{\text{c}} \widetilde{\mu}(\mu,\Gamma)}{\pi \hbar^2 v_{\text{F}}^2}
\end{gather}

\noindent with $A_{\text{c}}=5.24 \,\AA^2/(2\text{ atoms})$ being the elementary cell and $\rho(\mu,\Gamma)$ is measured in units of $\text{states}/\text{eV}\cdot \text{atom}$.

The intrinsic contribution disappears when spins are polarized perpendicular to the plane and the BR and ripple terms have a 2:1 anisotropy for the $\perp:\parallel$ directions. For the intrinsic part, $\Gamma_{\text{s},\text{i}}\approx \frac{L_{\text{i}}^2}{(2\mu)^2}\Gamma$ when $\mu \gg \Gamma$, which is an Elliott-Yafet like result with $\alpha_{\text{i}}=1$ since the band-band separation, $\Delta=2 \mu$. In the vicinity of the Dirac point, DP, (i.e. $\mu \approx 0$ and $\Gamma$ finite) it returns a Dyakonov-Perel like result of $\Gamma_{\text{s},\text{i}}=\frac{L_{\text{i}}^2}{4\ln(D/\Gamma)} \frac{1}{\Gamma}$. This is in agreement with the generalized Elliott-Yafet theory which predicts a similar crossover when the momentum scattering rate overcomes other energy scales \cite{DoraSimonStronglyCorrPRL2009}. Interestingly, the intrinsic contribution can be well fitted with a Lorentzian: $\Gamma_{\text{s},\text{i}} \approx \alpha' \frac{L_{\text{i}}^2\Gamma'}{\mu^2+\Gamma'^2}$, where $\alpha' \approx 0.2..0.4$ and $\Gamma'/\Gamma \approx 1..2$ for typical values of $\mu$ and $\Gamma$.

The BR term is only present if a perpendicular electric field, $E$, is applied, which induces a BR SOC of $L_{\text{BR}}=\kappa E$ with $\kappa$ values between 10 \cite{FabianPRB2009} and 36 $\mu\text{eV/(V/nm)}$ \cite{HuertasPRB2006}. The electric field changes $\mu$ through: $\mu=\sqrt{n \pi \hbar^2 v_{\text{F}}^2}$ where $n=\beta E$ is the carrier density and $\beta=0.22\,(\text{V}\cdot \text{nm})^{-1}$ for SiO$_2$ gate insulator \cite{GrapheneBias}. This yields the BR SOC as a function of $\mu$: $L_{\text{BR}}(\mu)\approx\kappa \mu^2 \cdot 3.4 \frac{\text{V}}{\text{eV}^2 \text{nm}}$.

The ripple relaxation contribution depends on $\Gamma$ only if $\mu \ll \Gamma$, where it resembles an EY relaxation: $\Gamma_{\text{s},\text{ripple}}\propto L^2_{\text{ripple}}\Gamma \ln(D/\Gamma)$.

\begin{figure}
\includegraphics[width=0.7\hsize]{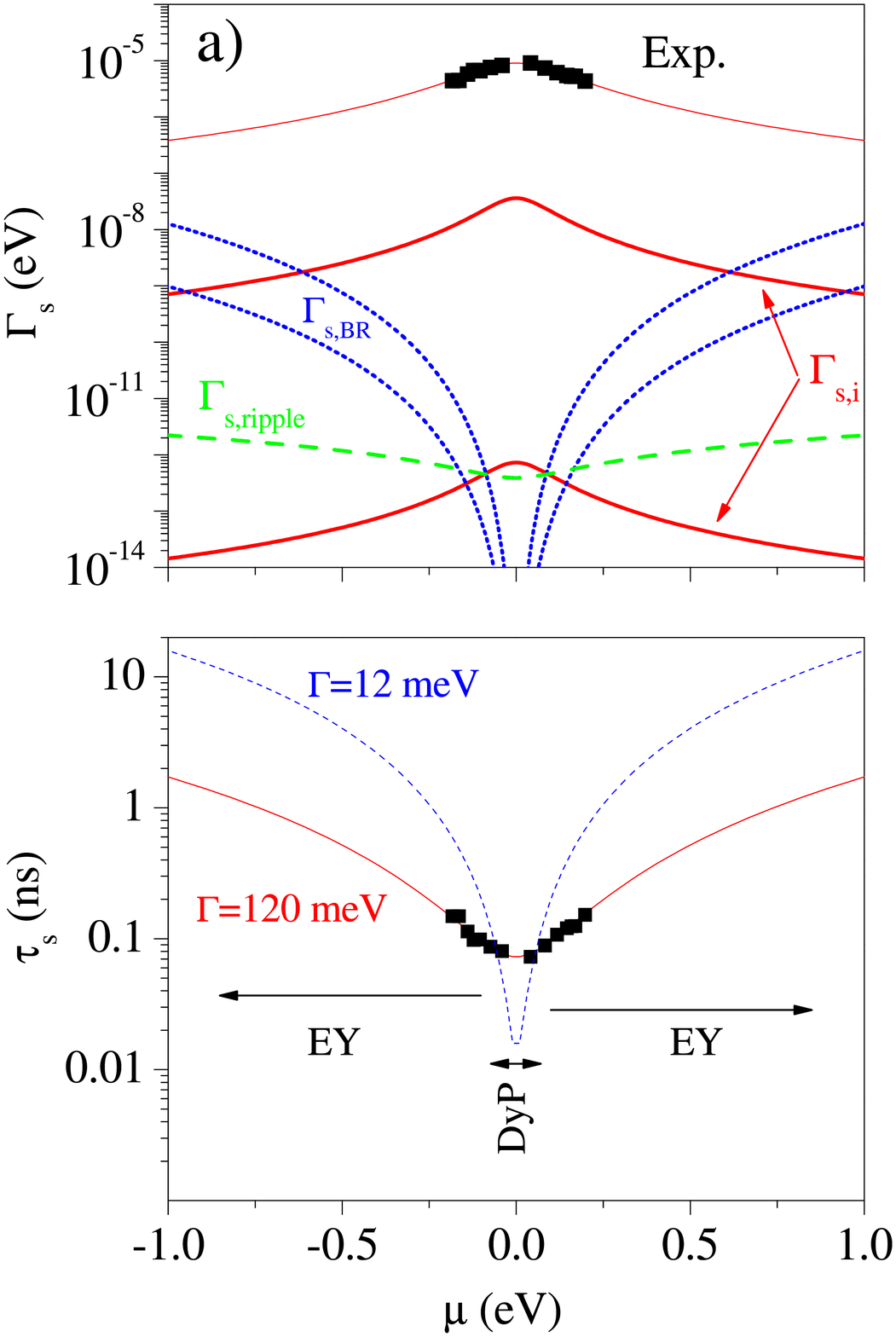}
\caption{a) Experimental (symbols, from Ref. \cite{JozsaPRB2009}) and calculated spin lattice relaxation rates, $\Gamma_{\text{s}}$, as a function of $\mu$ in graphene for in-plane spin polarization. Upper(lower) solid and dotted curves are the maximal(minimal) estimates for the intrinsic and BR contributions with SOC values from Refs. \cite{KaneMelePRL2005,HuertasPRB2006,FabianPRB2009}, respectively. Dashed curve is the ripple contribution with $L_{\text{ripple}}$ from Ref. \cite{HuertasPRB2006}. The upmost thin solid line is a fit to the data as explained in the text. b) The same experimental data shown as $\tau_{\text{s}}$ along with the fit (solid curve). For comparison, $\tau_{\text{s}}$ calculated with $\Gamma=12\,\text{meV}$ (dashed curve) is shown. Arrows depict the crossover of the DyP and EY mechanisms as a function of $\mu$.}
\label{Fig1_exp_vs_theory}
\end{figure}

\textit{Analysis of the spin transport data}. In the following, we analyze the available spin transport data \cite{JozsaNat,JozsaPRL2008,JozsaPRB2009} in the framework of the above calculation. Values of $\tau_{\text{s}}=60..125\,\text{ps}$ were found around the charge neutrality point (depending on the sample), with a typical $\Gamma \approx 75 \,\text{meV}$ \cite{JozsaPRB2009}. Fig. \ref{Fig1_exp_vs_theory}. shows the measured and calculated spin relaxation rate data for $\nu=\parallel$. $\Gamma=75\,\text{meV}$, that is independent of $\mu$, was used for the calculated curves. First principles calculations of the intrinsic SOC scatter more than two orders of magnitude with values of $0.9\,\mu\text{eV}$ \cite{GrapheneSOC_Min,GrapheneSOC_Yao,HuertasPRB2006}), $24\,\mu\text{eV}$ \cite{FabianPRB2009}, and $200\,\mu\text{eV}$ \cite{KaneMelePRL2005}. Values for the BR SOC, $L_{\text{BR}}=\kappa E$, vary between $\kappa=10..36\,\mu\text{eV/(V/nm)}$. This gives rise to the minimal and maximal estimates for both types of the contributions as shown in Fig. \ref{Fig1_exp_vs_theory}. The ripple SOC was estimated to be $17\,\mu\text{eV}$ in Ref. \cite{HuertasPRB2006}.

Clearly, the first principles based relaxation rates fall short of explaining the experimental observation. Of the three contributions, only
the intrinsic one has a $\mu$ dependence that mimics the experiment, whereas the other two shows the opposite. It may appear
that a fit to the data is ill defined, given the relatively large number of free parameters ($\Gamma$ and 3 $L$'s). However to our
surprise, the fit consistently yields the same, \emph{robust} set of parameters, irrespective of starting values or the method used
(least squares fitting or combined with a simulated annealing), which are: $L_\text{i}=3.7(1)\,\text{meV}$, $L_{\text{BR}}=L_{\text{ripple}}=0$, $\Gamma=120(5)\,\text{meV}$. This robustness originates from the qualitative difference between the $\mu$ dependence of the different contributions. The obtained values satisfy the criterion for the perturbative approach and the value of $\Gamma$ determined herein is in agreement with that obtained in Ref. \cite{JozsaPRB2009}.

The intrinsic SOC opens a bandgap of $L_{\text{i}}$ in the excitation spectrum \cite{KaneMelePRL2005,FabianPRB2009} therefore it is natural to ask: why is not this gap observed experimentally? Two interrelated answers are in order:
first, best quality samples to date are ballistic only on the (sub)micron scale, giving a momentum scattering rate of the order of meV's (or bigger),
which can mask the gap \cite{GeimRMP}. Second, charge inhomogeneities (the so-called puddles) prevent us from reaching the Dirac point, the average minimal charge density is estimated \cite{feldman} as $10^9$~cm$^{-2}$, which gives an average $\mu \sim 4\,\text{meV}$, capable of overwhelming the obtained gap.

The present analysis allows for the design of graphene based spintronic devices. For spins polarized perpendicular to the graphene plane, the intrinsic contribution vanishes thus resulting in a substantially longer spin relaxation time. For spins polarized in the graphene plane, Fig. \ref{Fig1_exp_vs_theory}. shows that around the Dirac point purer samples (i.e. smaller $\Gamma$) decreases $\tau_{\text{s}}$ rather
than increasing it, thus deteriorating performance. This, somewhat counterintuitive phenomenon, is the consequence of the Dyakonov-Perel
like behavior of the intrinsic contribution around the DP.

The large value obtained for the intrinsic SOC is surprising as it is an order of magnitude larger than the largest theoretical estimate \cite{KaneMelePRL2005} and up to 3 orders of magnitude larger than other results \cite{GrapheneSOC_Min,GrapheneSOC_Yao,HuertasPRB2006}). However, given that the experimental $\mu$ dependence of $\Gamma_{\text{s}}$ dictates the dominant role of the intrinsic coupling, $L_{\text{i}}$ yields necessarily a large value.
In the following, we consider two similar systems, MgB$_2$ and Li doped graphite and show that therein similar values of the intrinsic coupling are obtained.

In MgB$_2$, the boron atoms form a honeycomb lattice with four $p$-shell electrons, such as in graphene,
which highlights the similarity of the two materials. Therein, an intrinsic SOC of $L_{\text{i}}(\text{MgB}_2)=2.8\,\text{meV}$ of the $\pi$ orbitals was found \cite{SimonPRL2008}.
It was shown by Gr\"{u}neis and coworkers \cite{GrueneisPRB2009} and confirmed \cite{VallaCM} that alkali atom intercalated
graphite is an excellent model system of biased graphene as the two-dimensional electron dispersion is retained due to
the weak interlayer coupling. The Li intercalated stage I graphite compound LiC$_6$ \cite{DresselhausAP2002} is particularly suitable to determine
the intrinsic SOC as Li is the lightest alkali metal and its contribution to the spin relaxation is undetectable \cite{BeuneuMonodPRB1978}.

\begin{figure}
\includegraphics[width=0.7\hsize]{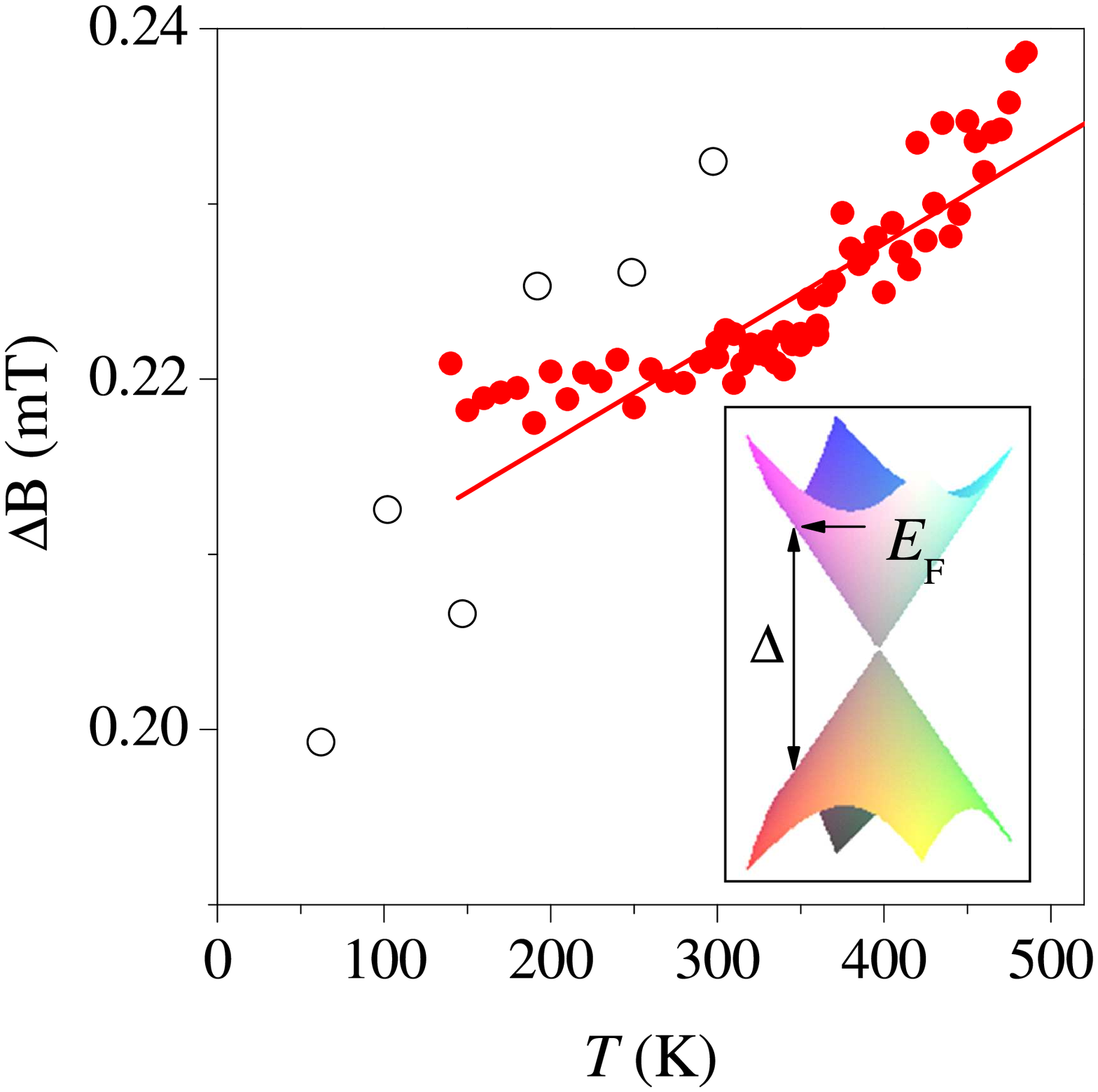}
\caption{High temperature ESR linewidth in HOPG LiC$_6$ (full symbols) and linear fit to the data (solid line). We show similar data from Ref. \cite{LauginiePhys1980} (open symbols) on a LiC$_6$ powder sample. Inset shows the schematics of the LiC$_6$ band structure according to the PES measurements \cite{GrueneisPRB2009,VallaCM} and the $\Delta=1.65\,\text{eV}$ parameter.}
\label{Fig2_Li_ESR}
\end{figure}

In Fig. \ref{Fig2_Li_ESR}., we show the temperature dependent ESR linewidth, $\Delta B$, for an HOPG LiC$_6$ along with previous data on a powder
LiC$_6$ sample \cite{LauginiePhys1980} and schematics of the band structure. A linear fit to the data yields
$\Delta B=0.205\,\text{[mT]}+T \times  6\cdot 10^{-5}\,\text{[mT/K]}$. Of these terms, the temperature dependent one is associated with the
homogeneous broadening, $\Delta B_{\text{hom}}$, due to SOC, which gives $\Gamma_{\text{s}}=g \mu_{\text{B}}\Delta B_{\text{hom}}$ ($g\approx 2$ is
the $g$-factor, $\mu_{\text{B}}$ is the Bohr magneton) and is $\Gamma_{\text{s}}=2.1\cdot 10^{-9}\,\text{eV}$ at 300 K. Since Li doped
graphite resembles biased graphene \cite{GrueneisPRB2009,VallaCM}, the above theory of the intrinsic SOC applies, i.e.
$\Gamma_{\text{s}}=\frac{L_{\text{i}}^2}{\Delta^2}\Gamma$. With the values of $\Delta=1.65\,\text{eV}$ \cite{VallaCM} and a
typical $\Gamma(300\,\text{K})=4.4\,\text{meV}$ \cite{DresselhausAP2002}, we obtain $L_{\text{i}}(\text{LiC}_6)=1.1\,\text{meV}$.
Although it is debated whether SOC in graphite is applicable for graphene \cite{HuertasPRB2006}, the similar result for these three systems leads us to conclude that the intrinsic SOC is properly determined in graphene.

\textit{Detectability of ESR on graphene}. With the SOC parameters and the theory of spin relaxation at hand, we assess the feasibility of ESR spectroscopy on graphene. It is determined by the sample amount, the magnitude of the spin-susceptibility, and the ESR linewidth. The ESR signal is proportional to the amount of magnetic moments: $\chi_0 V B/\mu_0$ where $B$ is the magnetic field, $\chi_0$ is the volume spin-susceptibility (dimensionless in SI units), $V$ is the sample volume, and $\mu_0$ is the permeability of vacuum. For graphene with area $A$, the amount of magnetic moments is $\chi_{0,\text{gr}} A B/\mu_0$ with $\chi_{0,\text{gr}}$ having a unit of meters. The Pauli spin-susceptibility of graphene is $\chi_{\text{0,gr}}=\mu_0  \mu_{\text{B}}^2 \rho(\mu,\Gamma)\frac{N}{A}$, $N$ is the number of carbons and the DOS, $\rho(\mu,\Gamma)$, is given above.

ESR spectrometer performance is given by the limit-of-detection (LOD$_0$) i.e. the number of $S=1/2$ non-interacting spins at 300 K which give a signal-to-noise of $S/N=10$ ratio for $\Delta B=0.1\,\text{mT}$ linewidth, and 1 sec/spectrum-point time constant. For state-of-the-art spectrometers $\text{LOD}_0 =10^{10} \text{ spins}/0.1$ mT. The spin-susceptibility of such spins is $\chi_{\text{Curie}}=\mu_0 \mu_{\text{B}}^2\frac{N_{\text{s}}}{V}/(k_{\text{B}}T)$, where $N_{\text{s}}$ spins occupy a volume of $V$, which gives an LOD for graphene:

\begin{gather}
\text{LOD}_{\text{gr}}= \text{LOD}_0 \cdot \frac{ f(\Delta B)}{26 \text{ meV}\times \rho\left(\mu,\Gamma\right) }
\end{gather}
in units of number of carbons. Here, 26 meV is the thermal energy at 300 K and the $f(\Delta B=0.1\,\text{mT})=1$ function describes that the ESR $S/N$ decreases as $1/\Delta B$ if $\Delta B<1$ mT (the typical magnetic field modulation limit) and as $1/(\Delta B)^2$ if $\Delta B\geq 1$ mT. For $\mu>\Gamma$, the DOS is well approximated by $\rho(\mu)=0.0385\mu\cdot[\text{states}/\text{eV}^2\text{atom}]$ which yields a compact result: $\text{LOD}_{\text{gr}}\approx 1000 \,\text{LOD}_0 \cdot f(\Delta B)/\mu$ with $\mu$ in eV units.

\begin{figure}
\includegraphics[width=0.8\hsize]{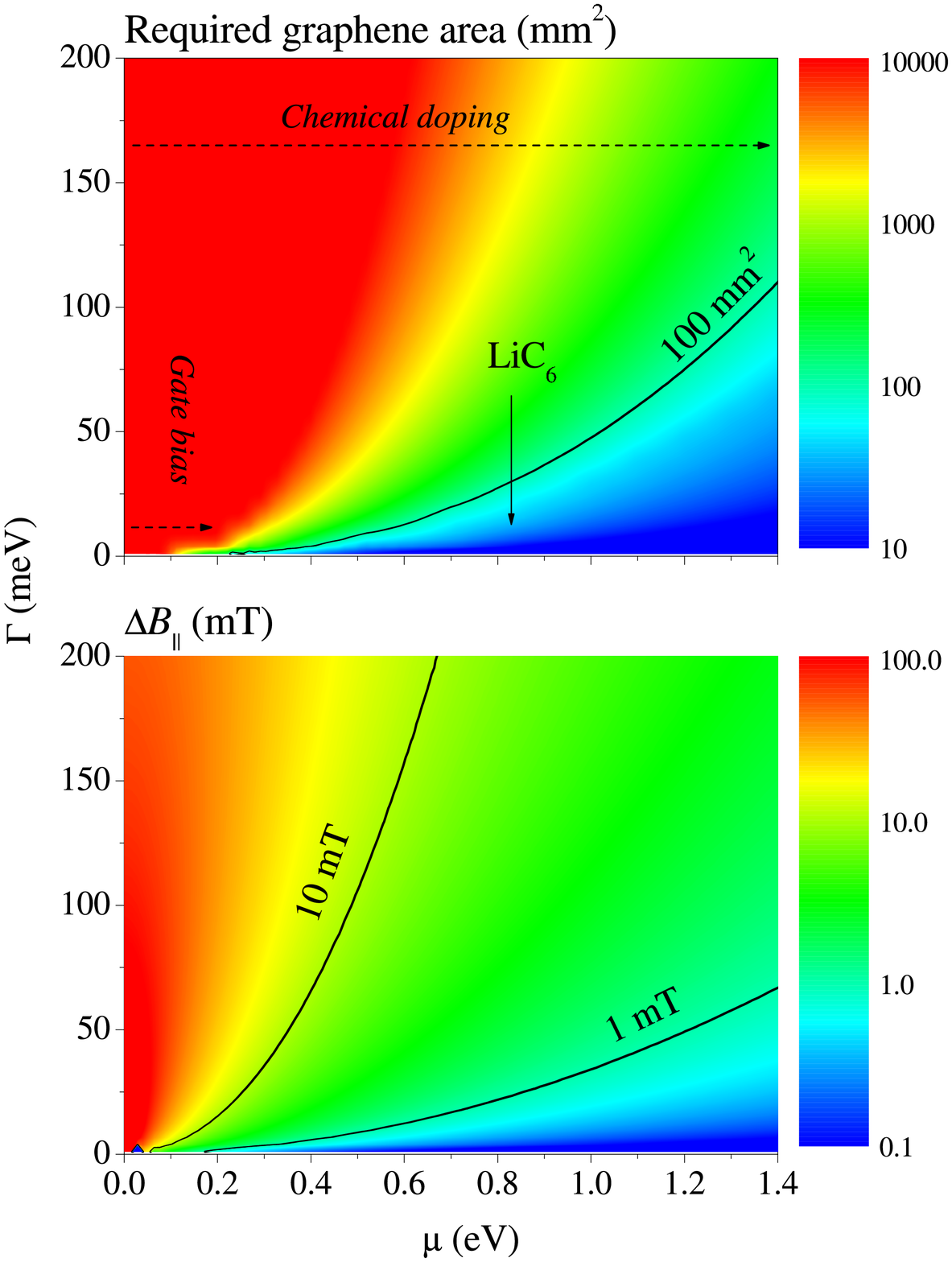}%
\caption{Limit of ESR detection for graphene as a function of $\mu$ and $\Gamma$ in units of the graphene area (upper panel) for an in-plane magnetic field. Arrows show the maximum chemical potential by gate bias and by chemical doping and solid curve indicate the 100 mm$^2$ area border. Expected ESR linewidth, $\Delta B_{\parallel}$ (lower panel), solid lines show two selected linewidths, 1 and 10 mT.}
\label{Fig3_ESR_sens}
\end{figure}

Clearly, a sizeable DOS and narrow linewidth are prerequisites to observe ESR on graphene. Large DOS can be achieved by moving $\mu$ away from the DP or by inducing defects.
The latter yields, however, increased scattering thus larger linewidth. Shifting $\mu$ by a gate bias is limited to $\sim 0.2\,\text{eV}$ due to breakdown in the most common SiO$_2$ insulator around $E\approx 0.1 \,\text{V/nm}$. With chemical doping using K, up to $\mu \sim 1.35\,\text{eV}$ can be achieved \cite{GrueneisPRB2009}. The ESR linewidth is expected to be strongly anisotropic with a minimum, $\Delta B_{\perp}$, for a perpendicular magnetic field, its magnitude however remains unknown. In Fig. \ref{Fig3_ESR_sens}., we show the calculated LOD for graphene as a function of $\mu$ and $\Gamma$ in units of the graphene area along with the calculated linewidth for an in-plane magnetic field. The LOD can be two orders of magnitude smaller for a perpendicular magnetic field if the corresponding ESR linewidth is an order of magnitude smaller. Therefore ESR experiments should be attempted with the perpendicular orientation first. This experiment would yield directly the magnitude of the BR and ripple relaxation contributions from $\Delta B_{\perp}$. 

An important benchmark, that indeed the intrinsic ESR signal of graphene is observed, is an angular dependence of the ESR linewidth:  $\Delta B(\theta)=\sin^2(\theta)\Delta B_{\|}+\cos^2(\theta)\Delta B_{\perp}$ as a function of the asimuth angle, $\theta$. Finally, we note that the anisotropy could reconcile the narrow ESR linewidth in the perpendicular geometry \cite{ForroPSSB2009} with the short $\tau_s$ in the spin transport experiment \cite{JozsaNat}.

In conclusion, we presented a theory of spin relaxation in graphene which takes into account intrinsic,
Bychkov-Rashba, and ripple spin-orbit coupling induced spin relaxation. Analysis of spin relaxation data show that the intrinsic contribution dominates the relaxation with a coupling constant that is orders
of magnitude larger than theoretical estimates but it is not unusually large compared to other honeycomb systems. The result predicts a large anisotropy of the spin relaxation. We presented under what circumstances bulk ESR spectroscopy can be observed in graphene.

Work supported by the Hungarian State Grants (OTKA) No. K72613. BD and FM acknowledge the Bolyai programme of the Hungarian Academy of Sciences and the Swiss National Foundation for support, respectively.



\begin{thebibliography}{31}
\expandafter\ifx\csname natexlab\endcsname\relax\def\natexlab#1{#1}\fi
\expandafter\ifx\csname bibnamefont\endcsname\relax
  \def\bibnamefont#1{#1}\fi
\expandafter\ifx\csname bibfnamefont\endcsname\relax
  \def\bibfnamefont#1{#1}\fi
\expandafter\ifx\csname citenamefont\endcsname\relax
  \def\citenamefont#1{#1}\fi
\expandafter\ifx\csname url\endcsname\relax
  \def\url#1{\texttt{#1}}\fi
\expandafter\ifx\csname urlprefix\endcsname\relax\def\urlprefix{URL }\fi
\providecommand{\bibinfo}[2]{#2}
\providecommand{\eprint}[2][]{\url{#2}}

\bibitem[{\citenamefont{Novoselov et~al.}(2004)\citenamefont{Novoselov, Geim,
  Morozov, Jiang, Zhang, Dubonos, Grigorieva, and Firsov}}]{NovoselovSCI2004}
\bibinfo{author}{\bibfnamefont{K.~S.} \bibnamefont{Novoselov}} \textit{et al.}, \bibinfo{journal}{Science}
  \textbf{\bibinfo{volume}{306}}, \bibinfo{pages}{666} (\bibinfo{year}{2004}).

\bibitem[{\citenamefont{\v{Z}uti\'c et~al.}(2004)\citenamefont{\v{Z}uti\'c,
  Fabian, and Sarma}}]{FabianRMP}
\bibinfo{author}{\bibfnamefont{I.}~\bibnamefont{\v{Z}uti\'c}},
  \bibinfo{author}{\bibfnamefont{J.}~\bibnamefont{Fabian}}, \bibnamefont{and}
  \bibinfo{author}{\bibfnamefont{S.~D.} \bibnamefont{Sarma}},
  \bibinfo{journal}{Rev. Mod. Phys.} \textbf{\bibinfo{volume}{76}},
  \bibinfo{pages}{323} (\bibinfo{year}{2004}).

\bibitem[{\citenamefont{Bolotin et~al.}(2008)\citenamefont{Bolotin, Sikes,
  Jiang, Klima, Fudenberg, Hone, Kim, and Stormer}}]{BolotinHighMobility}
\bibinfo{author}{\bibfnamefont{K.~I.} \bibnamefont{Bolotin}} \textit{et al.},
  \bibinfo{journal}{Sol. St. Comm.} \textbf{\bibinfo{volume}{146}},
  \bibinfo{pages}{351} (\bibinfo{year}{2008}).

\bibitem[{\citenamefont{Beuneu and Monod}(1978)}]{BeuneuMonodPRB1978}
\bibinfo{author}{\bibfnamefont{F.}~\bibnamefont{Beuneu}} \bibnamefont{and}
  \bibinfo{author}{\bibfnamefont{P.}~\bibnamefont{Monod}},
  \bibinfo{journal}{Phys. Rev. B} \textbf{\bibinfo{volume}{18}},
  \bibinfo{pages}{2422} (\bibinfo{year}{1978}).

\bibitem[{\citenamefont{Forr\'{o} et~al.}(1987)\citenamefont{Forr\'{o},
  Sekretarczyk, Krupski, Schweitzer, and Keller}}]{ForroPRB1987}
\bibinfo{author}{\bibfnamefont{L.}~\bibnamefont{Forr\'{o}}} \textit{et al.},
  \bibinfo{journal}{Phys. Rev. B} \textbf{\bibinfo{volume}{35}},
  \bibinfo{pages}{2501} (\bibinfo{year}{1987}).

\bibitem[{\citenamefont{Tombros et~al.}(2007)\citenamefont{Tombros, J\'{o}zsa,
  Popinciuc, Jonkman, and van Wees}}]{JozsaNat}
\bibinfo{author}{\bibfnamefont{N.}~\bibnamefont{Tombros}} \textit{et al.}, \bibinfo{journal}{Nature} \textbf{\bibinfo{volume}{448}},
  \bibinfo{pages}{571} (\bibinfo{year}{2007}).

\bibitem[{\citenamefont{Tombros et~al.}(2008)\citenamefont{Tombros, Tanabe,
  Veligura, J\'{o}zsa, Popinciuc, Jonkman, and van Wees}}]{JozsaPRL2008}
\bibinfo{author}{\bibfnamefont{N.}~\bibnamefont{Tombros}} \textit{et al.}, \bibinfo{journal}{Phys. Rev. Lett.} \textbf{\bibinfo{volume}{101}},
  \bibinfo{pages}{046601} (\bibinfo{year}{2008}).

\bibitem[{\citenamefont{Elliott}(1954)}]{Elliott}
\bibinfo{author}{\bibfnamefont{R.~J.} \bibnamefont{Elliott}},
  \bibinfo{journal}{Phys. Rev.} \textbf{\bibinfo{volume}{96}},
  \bibinfo{pages}{266–} (\bibinfo{year}{1954}).

\bibitem[{\citenamefont{Yafet}(1963)}]{YafetReview}
\bibinfo{author}{\bibfnamefont{Y.}~\bibnamefont{Yafet}},
  \bibinfo{journal}{Solid State Phys.} \textbf{\bibinfo{volume}{14}},
  \bibinfo{pages}{1} (\bibinfo{year}{1963}).

\bibitem[{\citenamefont{Dresselhaus}(1955)}]{DresselhausPR1955}
\bibinfo{author}{\bibfnamefont{G.}~\bibnamefont{Dresselhaus}},
  \bibinfo{journal}{Phys. Rev.} \textbf{\bibinfo{volume}{100}},
  \bibinfo{pages}{580} (\bibinfo{year}{1955}).

\bibitem[{\citenamefont{Bychkov and
  Rashba}(1984{\natexlab{a}})}]{BychkovRashba1}
\bibinfo{author}{\bibfnamefont{Y.~A.} \bibnamefont{Bychkov}} \bibnamefont{and}
  \bibinfo{author}{\bibfnamefont{E.~I.} \bibnamefont{Rashba}},
  \bibinfo{journal}{J. Phys. C} \textbf{\bibinfo{volume}{17}},
  \bibinfo{pages}{6039} (\bibinfo{year}{1984}{\natexlab{a}}).

\bibitem[{\citenamefont{Bychkov and
  Rashba}(1984{\natexlab{b}})}]{BychkovRashba2}
\bibinfo{author}{\bibfnamefont{Y.~A.} \bibnamefont{Bychkov}} \bibnamefont{and}
  \bibinfo{author}{\bibfnamefont{E.~I.} \bibnamefont{Rashba}},
  \bibinfo{journal}{JETP Lett.} \textbf{\bibinfo{volume}{39}},
  \bibinfo{pages}{78} (\bibinfo{year}{1984}{\natexlab{b}}).

\bibitem[{\citenamefont{D\'{o}ra and
  Simon}(2009)}]{DoraSimonStronglyCorrPRL2009}
\bibinfo{author}{\bibfnamefont{B.}~\bibnamefont{D\'{o}ra}} \bibnamefont{and}
  \bibinfo{author}{\bibfnamefont{F.}~\bibnamefont{Simon}},
  \bibinfo{journal}{Phys. Rev. Lett.} \textbf{\bibinfo{volume}{102}},
  \bibinfo{pages}{137001} (\bibinfo{year}{2009}).

\bibitem[{\citenamefont{Huertas-Hernando
  et~al.}(2006)\citenamefont{Huertas-Hernando, Guinea, and
  Brataas}}]{HuertasPRB2006}
\bibinfo{author}{\bibfnamefont{D.}~\bibnamefont{Huertas-Hernando}},
  \bibinfo{author}{\bibfnamefont{F.}~\bibnamefont{Guinea}}, \bibnamefont{and}
  \bibinfo{author}{\bibfnamefont{A.}~\bibnamefont{Brataas}},
  \bibinfo{journal}{Phys. Rev. B} \textbf{\bibinfo{volume}{74}},
  \bibinfo{pages}{155426} (\bibinfo{year}{2006}).

\bibitem[{\citenamefont{Gmitra et~al.}(2009)\citenamefont{Gmitra, Konschuh,
  Ertler, Ambrosch-Draxl, and Fabian}}]{FabianPRB2009}
\bibinfo{author}{\bibfnamefont{M.}~\bibnamefont{Gmitra}} \textit{et al.},
  \bibinfo{journal}{Phys. Rev. B} \textbf{\bibinfo{volume}{80}},
  \bibinfo{pages}{235431} (\bibinfo{year}{2009}).

\bibitem[{\citenamefont{Kane and Mele}(2005)}]{KaneMelePRL2005}
\bibinfo{author}{\bibfnamefont{C.~L.} \bibnamefont{Kane}} \bibnamefont{and}
  \bibinfo{author}{\bibfnamefont{E.~J.} \bibnamefont{Mele}},
  \bibinfo{journal}{Phys. Rev. Lett.} \textbf{\bibinfo{volume}{95}},
  \bibinfo{pages}{226801} (\bibinfo{year}{2005}).

\bibitem[{\citenamefont{Ertler et~al.}()\citenamefont{Ertler, Konschuh, Gmitra,
  and Fabian}}]{FabianCM0905.0424}
\bibinfo{author}{\bibfnamefont{C.}~\bibnamefont{Ertler}} \textit{et al.}, \bibinfo{howpublished}{arXiv:0905.0424v2}.

\bibitem[{\citenamefont{J\'{o}zsa et~al.}(2009)\citenamefont{J\'{o}zsa,
  Maassen, Popinciuc, Zomer, Veligura, Jonkman, and van Wees}}]{JozsaPRB2009}
\bibinfo{author}{\bibfnamefont{C.}~\bibnamefont{J\'{o}zsa}} \textit{et al.}, \bibinfo{journal}{Phys. Rev. B} \textbf{\bibinfo{volume}{80}},
  \bibinfo{pages}{241403(R)} (\bibinfo{year}{2009}).

\bibitem[{\citenamefont{Griswold et~al.}(1952)\citenamefont{Griswold, Kip, and
  Kittel}}]{KipKittelPR1952}
\bibinfo{author}{\bibfnamefont{T.~W.} \bibnamefont{Griswold}},
  \bibinfo{author}{\bibfnamefont{A.~F.} \bibnamefont{Kip}}, \bibnamefont{and}
  \bibinfo{author}{\bibfnamefont{C.}~\bibnamefont{Kittel}},
  \bibinfo{journal}{Physical Review} \textbf{\bibinfo{volume}{88}},
  \bibinfo{pages}{951} (\bibinfo{year}{1952}).

\bibitem[{\citenamefont{Ciric et~al.}(2009)\citenamefont{Ciric, Sienkiewicz,
  N\'{a}fr\'{a}di, Mionic, Magrez, and Forr\'{o}}}]{ForroPSSB2009}
\bibinfo{author}{\bibfnamefont{L.}~\bibnamefont{Ciric}} \textit{et al.},
  \bibinfo{journal}{Phys. Stat. Sol. B} \textbf{\bibinfo{volume}{246}},
  \bibinfo{pages}{2558} (\bibinfo{year}{2009}).

\bibitem[{\citenamefont{Zanini et~al.}(1978)\citenamefont{Zanini, Basu, and
  Fischer}}]{FischerLiC6}
\bibinfo{author}{\bibfnamefont{M.}~\bibnamefont{Zanini}},
  \bibinfo{author}{\bibfnamefont{S.}~\bibnamefont{Basu}}, \bibnamefont{and}
  \bibinfo{author}{\bibfnamefont{J.~E.} \bibnamefont{Fischer}},
  \bibinfo{journal}{Carbon} \textbf{\bibinfo{volume}{16}}, \bibinfo{pages}{211}
  (\bibinfo{year}{1978}).

\bibitem[{\citenamefont{Dresselhaus and Dresselhaus}(2002)}]{DresselhausAP2002}
\bibinfo{author}{\bibfnamefont{M.~S.} \bibnamefont{Dresselhaus}}
  \bibnamefont{and}
  \bibinfo{author}{\bibfnamefont{G.}~\bibnamefont{Dresselhaus}},
  \bibinfo{journal}{Adv. Phys.} \textbf{\bibinfo{volume}{51}},
  \bibinfo{pages}{1} (\bibinfo{year}{2002}).

\bibitem[{\citenamefont{Simon et~al.}(2008)\citenamefont{Simon, D\'{o}ra,
  Mur\'{a}nyi, J\'{a}nossy, Garaj, Forr\'{o}, Bud'ko, Petrovic, and
  Canfield}}]{SimonPRL2008}
\bibinfo{author}{\bibfnamefont{F.}~\bibnamefont{Simon}} \textit{et al.}, \bibinfo{journal}{Phys. Rev. Lett.} \textbf{\bibinfo{volume}{101}},
  \bibinfo{pages}{177003} (\bibinfo{year}{2008}).

\bibitem[{\citenamefont{Fern\'{a}ndez-Rossier
  et~al.}(2007)\citenamefont{Fern\'{a}ndez-Rossier, Palacios, and
  Brey}}]{GrapheneBias}
\bibinfo{author}{\bibfnamefont{J.}~\bibnamefont{Fern\'{a}ndez-Rossier}},
  \bibinfo{author}{\bibfnamefont{J.~J.} \bibnamefont{Palacios}},
  \bibnamefont{and} \bibinfo{author}{\bibfnamefont{L.}~\bibnamefont{Brey}},
  \bibinfo{journal}{Phys. Rev. B} \textbf{\bibinfo{volume}{75}},
  \bibinfo{pages}{205441} (\bibinfo{year}{2007}).

\bibitem[{\citenamefont{Min et~al.}(2006)\citenamefont{Min, Hill, Sinitsyn,
  Sahu, Kleinman, and MacDonald}}]{GrapheneSOC_Min}
\bibinfo{author}{\bibfnamefont{H.}~\bibnamefont{Min}} \textit{et al.}, \bibinfo{journal}{Phys. Rev. B} \textbf{\bibinfo{volume}{77}},
  \bibinfo{pages}{165310} (\bibinfo{year}{2006}).

\bibitem[{\citenamefont{Yao et~al.}(2007)\citenamefont{Yao, Ye, Qi, Zhang, and
  Fang}}]{GrapheneSOC_Yao}
\bibinfo{author}{\bibfnamefont{Y.}~\bibnamefont{Yao}} \textit{et al.}, \bibinfo{journal}{Phys. Rev. B} \textbf{\bibinfo{volume}{75}},
  \bibinfo{pages}{041401} (\bibinfo{year}{2007}).

\bibitem[{\citenamefont{Castro~Neto et~al.}(2009)\citenamefont{Castro~Neto,
  Guinea, Peres, Novoselov, and Geim}}]{GeimRMP}
\bibinfo{author}{\bibfnamefont{A.~H.} \bibnamefont{Castro~Neto}} \textit{et al.}, \bibinfo{journal}{Rev. Mod. Phys.} \textbf{\bibinfo{volume}{81}},
  \bibinfo{pages}{109} (\bibinfo{year}{2009}).

\bibitem[{\citenamefont{Feldman et~al.}(2009)\citenamefont{Feldman, Martin, and
  Yacoby}}]{feldman}
\bibinfo{author}{\bibfnamefont{B.~E.} \bibnamefont{Feldman}},
  \bibinfo{author}{\bibfnamefont{J.}~\bibnamefont{Martin}}, \bibnamefont{and}
  \bibinfo{author}{\bibfnamefont{A.}~\bibnamefont{Yacoby}},
  \bibinfo{journal}{Nature Physics} \textbf{\bibinfo{volume}{5}},
  \bibinfo{pages}{889} (\bibinfo{year}{2009}).

\bibitem[{\citenamefont{Gr\"{u}neis et~al.}(2009)\citenamefont{Gr\"{u}neis,
  Attaccalite, Rubio, Vyalikh, Molodtsov, Fink, Follath, Eberhardt,
  B\"{u}chner, and Pichler}}]{GrueneisPRB2009}
\bibinfo{author}{\bibfnamefont{A.}~\bibnamefont{Gr\"{u}neis}} \textit{et al.}, \bibinfo{journal}{Phys. Rev. B} \textbf{\bibinfo{volume}{80}},
  \bibinfo{pages}{075431} (\bibinfo{year}{2009}).

\bibitem[{\citenamefont{Pan et~al.}()\citenamefont{Pan, Camacho, Upton,
  Fedorov, Walters, Howard, Ellerby, and Valla}}]{VallaCM}
\bibinfo{author}{\bibfnamefont{Z.-H.} \bibnamefont{Pan}} \textit{et al.}, \bibinfo{howpublished}{arXiv:1003.3903}.

\bibitem[{\citenamefont{Lauginie et~al.}(1980)\citenamefont{Lauginie, Estrade,
  Conard, Gu\'{e}rard, Lagrange, and El~Makrini}}]{LauginiePhys1980}
\bibinfo{author}{\bibfnamefont{P.}~\bibnamefont{Lauginie}} \textit{et al.}, \bibinfo{journal}{Physica B} \textbf{\bibinfo{volume}{99}},
  \bibinfo{pages}{514} (\bibinfo{year}{1980}).

\end{thebibliography}

\end{document}